\documentclass[12pt,a4paper]{article}

\usepackage{indentfirst}
\usepackage{rotating}
\usepackage{graphicx}

\begin{document}


\title{\Large \bf Interference effects in hyperfine induced $2s2s~^3P_0,~^3P_2 \rightarrow 2s^2~^1S_0$ transitions of Be-like ions}

\author{{\small \rm Jiguang Li and Chenzhong Dong \footnote{Tel. No.: +86(0)931 7971148, Fax No.: +86(0)931 7971277, E-mail: Dongcz@nwnu.edu.cn}}\\
\footnotesize  $^{\dagger}$ College of Physics and Electronic Engineering, Northwest Normal University, Lanzhou 730070, China \\
\footnotesize  $^{\ddag}$ Joint Laboratory of Atomic and Molecular Physics, NWNU \& IMPCAS, Lanzhou 730070, China \\
}
\date{}

\maketitle

\begin{abstract}
The hyperfine induced $2s2p~^3P_0, ~^3P_2 \rightarrow 2s^2~^1S_0$ E1
transition probabilities of Be-like ions were calculated using
grasp2K based on multi-configuration Dirac-Fock method and HFST
packages. It was found that the hyperfine quenching rates are
strongly affected by the interference for low-Z Be-like ions,
especially for $2s2p~^3P_0 \rightarrow 2s^2~^1S_0$ transition. In
particular, the trends of interference effects with atomic number $Z$
in such two transitions are not monotone. The strongest interference
effect occurs near $Z=7$ for $2s2p~^3P_0 \rightarrow 2s^2~^1S_0$ E1
transition, and near $Z=9$ for $2s2p~^3P_2 \rightarrow 2s^2~^1S_0$
E1 transition.
\end{abstract}


{\small  key words: Hyperfine induced transition; Hyperfine spectroscopy; Interference effect; Be-like ions.}

\section{Introduction}
Hyperfine spectroscopy is very important tool in study of atomic and
nuclear physics, especially which can be used to check fundamental
interaction \cite{Bennett, Bouchiat, Labzowsky, Bondarevskaya, Li}
such as electromagnetic and electroweak interaction with high
accuracy, determine nuclear properties \cite{Bieron, Beloy, Okada,
Fm, Labzowsky2}, develop atomic clock \cite{Becker, Takamoto,
Petersen, Porsev, Joensson}, and so on. Recently one kind of
hyperfine transitions, which is known as hyperfine induced
transition or hyperfine quenching, attracts more attention owing to
analyzing spectra \cite{Andersson1}, determining isotopic ratios in
stellar and diagnosing low-density plasma \cite{Brage,Brage2}
besides those reasons mentioned above.

Many theoretical researches have been made to predict hyperfine
induced transition probability based on different method. The most
important problem in the calculations of hyperfine quenching rate is
how to treat the hyperfine interaction and the interaction with
electromagnetic field. In the past decades, three methods have been
developed to solve such problem, i.e. perturbative method
\cite{Marrus, Aboussaid}, complex matrix method
\cite{Indelicato,Indelicato2} and radiative damping method
\cite{Johnson,Cheng}. They have individual merit and shortage in
actual physical problems as shown in Ref. \cite{Johnson,Cheng}. Here
we used perturbative method to evaluate hyperfine quenching rate,
that is, the computations of the probability follow the
determinations of hyperfine structure. This approach is valid as
long as radiative widths are smaller than the fine separation
between concerned levels. The crucial point is to properly choice
perturbative states involved in because the differences of results
calculated by different approximate model are sometimes very large.

For hyperfine induced $2s2p~^3P_0 \rightarrow 2s^2~^1S_0$ transition
of Be-like ions there have been several studies. Marques {\it et
al.} firstly computed the probabilities using their developing
complex matrix method \cite{Marques}. However, in their
computational model they neglected an important contribution from
the $2s2p~^1P_1$ level, which lead to a relative large discrepancy
compared to the later theoretical \cite{Brage, Cheng} and
experimental value \cite{Brage2, Schippers}. Once again, Brage {\it
et al.} gave some of transition rates by means of perturbative
method in order to determine the isotopic of composition and
diagnose densities of low-density plasmas\cite{Brage}. In their
investigation the influence of $2s2p~^1P_1$ level on hyperfine
induced $2s2p~^3P_0 \rightarrow 2s^2~^1S_0$ transition of Be-like
ions were indicated. But the calculations were restricted within
those abundant elements in stellar. Later, Schippers {\it et al.}
measured the probability for Be-like Ti ions using resonant
electron-ion recombination method in the heavy-ion storage ring TSR
of Max-Planck Institute for Nuclear Physics, Heidelberg, Germany
\cite{Schippers}. The experimental result was almost 60\% larger
than Marques {\it et al.} theoretical value. This led to Cheng {\it
et al.} renewedly calculating this decay rate along Be-like
isoelectronic sequence by perturbative and radiative damping method
\cite{Cheng}. The latest theoretical results reduced the discrepancy
to 20\%. Their investigations shown again that the contribution from
$2s2p~^1P_1$ transition amplitude and the interference effect from
$2s2p~3P_1$ and $2s2p~^1P_1$ perturbative states on hyperfine
induced $2s2s~3P_0 \rightarrow 2s^2~^1S_0$ transition of Be-like
ions can not be neglected. Even though, it seems that characteristic
about interference effects in hyperfine induced $2s2p~^3P_0
\rightarrow 2s^2~^1S_0$ transition of Be-like ions were still not
very clear to be revealed.

As we know, $^3P_2$ level is another metastable state of sp
configuration. It can decay to the excited state $^3P_1$ by magnetic
dipole (M1) transition and to the ground state by magnetic
quadurpole (M2) transition with large branch ratio. Many
publications have been concerned with determination of the
probabilities \cite{Trabert, Kingston, Tupitsyn, Glass, Majumder} to
diagnose the low-density plasma and to probe relativistic and QED
effects by accurate transition energy and probability. However,
information about another transition process, that is, hyperfine
induced E1 transition is very scarce. While, Gould {\it et al.} and
Andersson {\it et al.} have pointed out significant influences of
hyperfine induced E1 transition on the lifetime of $^3P_2$ level in
He-like \cite{Gould} and Zn-like ions \cite{Andersson2},
respectively. Dubau {\it et al.} have also shown that quantum
interference between the E1 transition of hyperfine induced and M2
transition has obvious effects on increasing the degree of linear
polarization of $^3P_2 \rightarrow~^1S_0$ in He-like ions
\cite{Dubau}, which can affect the modelling and diagnostics of
high-temperature astrophysical and laboratory plasmas with an
anisotropic non-Maxwellian velocity distribution of energetic
electrons.

Based on these reasons mentioned above, we further investigated
hyperfine induced $2s2p~^3P_2, ~^3P_0 \rightarrow 2s^2~^1S_0$
transition of Be-like ions in detail using grasp2K \cite{grasp2K}
based on multi-configuration Dirac-Fock and HFST \cite{HFST}
packages. The regular of interference effects for hyperfine induced
$2s2s~^3P_0,~^3P_2 \rightarrow 2s^2~^1S_0$ transition of Be-like
ions was shown. In particular, it was found that the trends of interference effects
with atomic number $Z$ in such two transitions are not monotone.

\section{Theory}
\subsection{Wavefunction of hyperfine level}
In the present of hyperfine interactions, which couple the nuclear
{\bf $I$} and electronic {\bf $J$} angular momenta to a total
angular momenta {\bf $F=I+J$}, only {\bf $F$} and $M_{F}$ are good
quantum number other than the electronic angular momenta {\bf $J$}
and $M_J$. Then, the wave function for the system can be written by
\begin{eqnarray}
|F M_F \rangle = \sum_{\gamma J} h_{\gamma J} |\gamma J I F
M_F\rangle,
\end{eqnarray}
where $h_{\gamma, J}$ is the mixing coefficients due to hyperfine
interaction, and are obtained in first order perturbation theory as
the ratio between the off-diagonal hyperfine matrix elements and the
unperturbed energy differences
\begin{equation}
h_{\gamma J} = \frac{\langle \gamma J I F M_F | H_{hfs} |\gamma_0
J_0 I F M_F \rangle}{E(\gamma_0 J_0) - E(\gamma J)},
\end{equation}
the subscript 0 labels the concerned level. The hyperfine
interaction Hamiltonian $H_{hfs}$ in this formula can be represented
as a multipole expansion \cite{HFS92, HFSZEEMAN},
\begin{eqnarray}
H_{hfs} = \sum_{k \ge 1} {\bf T^{k}} \cdot {\bf M^{k}},
\end{eqnarray}
where ${\bf T^{(k)}}$ and ${\bf M^{(k)}}$ are spherical tensor
operators of rank k in the electronic and nuclear spaces,
respectively \cite{HFS92}. In the following discussion we only
include main the nuclear magnetic dipole (k=1) and electric
quadrupole (k=2) interaction. Applying Racah's algebra, hyperfine
interaction matrix elements can be further written by \cite{Edmonds}
{\footnotesize
\begin{eqnarray}
\langle \gamma J I F M_F | H_{hfs} |\gamma_0 J_0 I F M_F \rangle =
(-1)^{I+J_0-F} [(2J+1)(2I+1)]^{1/2}\left \{
                                            \begin{array}{ccc}
                                               I & J & F \\
                                               J_0& I & 1
                                            \end{array}
                                    \right \} \langle \gamma J || {\bf T^{(1)}} || \gamma J_0 \rangle \langle I || {\bf M^{(1)}} || I \rangle, \\
\langle \gamma J I F M_F | H_{hfs} |\gamma_0 J_0 I F M_F \rangle =
(-1)^{I+J_0-F} [(2J+1)(2I+1)]^{1/2}\left \{
                                            \begin{array}{ccc}
                                               I & J & F \\
                                               J_0& I & 2
                                            \end{array}
                                    \right \} \langle \gamma J || {\bf T^{(2)}} || \gamma J_0 \rangle \langle I || {\bf M^{(2)}} || I \rangle.
\end{eqnarray}
}

The reduce matrix elements of the tensor {$\bf M^{(k)}$} are related
to the conventionally defined nuclear moment,
\begin{eqnarray}
\langle I || {\bf M^{(1)}} || I \rangle &=& \mu_I \sqrt{\frac{I+1}{I}}, \\
\langle I || {\bf M^{(2)}} || I \rangle &=& \frac{Q}{2}
\sqrt{\frac{(2I+3)(I+1)}{I(2I-1)}}.
\end{eqnarray}
where $\mu_I$ is nuclear magnetic dipole moment in $\mu_N$ of the
nuclear magneton, and $Q$ is electric quadrupole moment in barns.

\subsection{Hyperfine transition probability}
The electric dipole (E1) transition probability between two
different hyperfine levels $|F M_F \rangle$ and $|F' M'_F \rangle$
is given by \cite{Cowan}
\begin{eqnarray}
A = \frac{4 {\omega}^3}{3 c^3} \sum_{M_F} |\langle F M_F | {\bf
Q^{(1)}} | F' M'_F \rangle|^2,
\end{eqnarray}
where {$\bf Q^{(1)}$} is the electric dipole tensor operator.
Substitute (1) into above formula, then
\begin{eqnarray}
A = \frac{4 {\omega}^3}{3 c^3} \frac{1}{2F'+1} |\sum_{\gamma J} \sum_{\gamma' J'} h_{\gamma J} h_{\gamma' J'} \langle \gamma J I F || Q^{(1)} || \gamma' J' I F' \rangle|^2.
\end{eqnarray}
Because operator $Q^{(1)}$ only act on electronic parts, reduced
matrix element $\langle \gamma J I F ||  Q^{(1)} || \gamma' J'
I F' \rangle$ can be simplified based on Racah's algbra
\cite{Edmonds},
\begin{eqnarray}
\langle \gamma J I F ||  Q^{(1)} || \gamma' J' I F' \rangle &=
(-1)^{(J+I+F')} \sqrt{(2F+1)(2F'+1)} \left \{
        \begin{array}{ccc}
             J & F & I \\
             F'& J' & 1
        \end{array}
\right \} \langle \gamma J ||  Q^{(1)} || \gamma' J' \rangle, \nonumber \\
\end{eqnarray}
therefore,
\begin{eqnarray}
A &=& \frac{4 {\omega}^3}{3 c^3} (2F+1) |\sum_{\gamma J}
\sum_{\gamma' J'} h_{\gamma J} h_{\gamma' J'} \left \{
        \begin{array}{ccc}
             J & F & I \\
             F'& J' & 1
        \end{array}
\right \} \langle \gamma J ||  Q^{(1)} || \gamma' J' \rangle|^2, \nonumber \\
\end{eqnarray}
where $\omega$ is the transition energy in Hartree. The reduced
transition matrix elements of the electric dipole operator can be
obtained as square roots of the corresponding line strengths.

Using similar method one can obtain other type hyperfine induced
transition probability such as M1, E2, etc. \cite{Li, Johnson,
Andersson3}.

As can be seen from the derivation, hyperfine transition probability
depends on nuclear parameters. It is not convenience for us to
further discuss the trend of the rate along atomic number $Z$.
Therefore, we generalized Brage {\it et al.} method \cite{Brage} so
that hyperfine transition rate is independent of nuclear properties,
which is called reduced hyperfine transition probability A$^{el}$.
By defining reduced hyperfine mixing coefficient $h^{el}$,
\begin{eqnarray}
h^{el} = \frac{(-1)^{-(I+J_0+F)} h}{\mu_I
[(1+I^{-1})(2I+1)]^{1/2}W(IJ_0\ JI;F1)},
\end{eqnarray}
then,
\begin{eqnarray}
A^{el} &=& \frac{4 {\omega}^3}{3 c^3} (2F+1) |\sum_{\gamma J}
\sum_{\gamma' J'} h^{el}_{\gamma J} h^{el}_{\gamma' J'} \left \{
        \begin{array}{ccc}
             J & F & I \\
             F'& J' & 1
        \end{array}
\right \} \langle \gamma J || Q^{(1)} || \gamma' J' \rangle|^2 \nonumber \\
\end{eqnarray}
where $W(IJ_0\ JI;F1)$ are 6j-symbol in eq(4). To simplify we
neglected electric quadrupole hyperfine interaction in above
equation due to quite weak compared to the magnetic dipole
interaction.

\subsection{Electronic wave function}
The electronic wave functions $|\gamma J \rangle$ were computed
using the grasp2K program package \cite{grasp2K}. Here the wave
function for a state labeled $\gamma J$ is approximated by an
expansion over $jj$-coupled configuration state functions (CSFs)
\begin{equation}
|\gamma J \rangle = \sum_j c_j \Phi_j.
\end{equation}
The configuration state functions $\Phi_j$ are anti-symmetrized
linear combinations of products of Dirac orbitals. In the
multi-configuration self-consistent field (SCF) procedure both the
radial parts of the orbitals and the expansion coefficients are
optimized to self-consistency. In the present work a Dirac-Coulomb
Hamiltonian was used with the nucleus described by an extended Fermi
charge distribution.

Once the radial orbitals have been determined relativistic
configuration interaction (CI) calculations can be performed. Here
higher-order interactions may be included in the Hamiltonian. The
most important of these is the Breit interactions
\begin{eqnarray}
H_{Breit} &=& - \sum_{i<j}^N [\frac{{\bf \alpha}_{i} \cdot {\bf
\alpha}_{j} \cos(\omega_{ij} r_{ij})}{r_{ij}}
          + ({\bf \alpha}_{i} \cdot \nabla_i)({\bf \alpha}_{j} \cdot \nabla_j) \frac{\cos(\omega_{ij} r_{ij}) - 1}{\omega_{ij}^2 r_{ij}}] ,
\end{eqnarray}
where photon frequency $\omega_{ij}$ is obtained as the difference
between the diagonal Lagrange multipliers $\epsilon_i$ and
$\epsilon_j$ associated with the orbitals. However, this is invalid
when shells are multiply occupied, and the diagonal energy
parameters of correlation orbitals with small occupation numbers may
be large positive quantities totally unrelated to binding energies
\cite{Fischer, Fischerbook}. For this reason, the zero-frequency
limit have been adopted in present calculations. In the
configuration interaction calculations the main quantum
electrodynamics (QED) effects can also be included.

Tensor algebra used for evaluating hyperfine and electric dipole
matrix elements between CI wave functions assumes that the wave
functions are built from a common orbital set. This is a severe
restriction since high-quality wave functions demands orbitals
optimized for the specific state. To relax this and to be able to
compute matrix elements between wave functions built from
independently optimized orbital sets, biorthogonal transformation
techniques introduced by Malmqvist can be used \cite{Malmqvist,
Biotra}.

\section{Results and discussions}
\subsection{Calculational model and method}
The accuracy of the calculated hyperfine induced transition rate
depends on the number of perturbative states in Eq. (11) and on the accuracy of
the electronic matrix elements. In practical calculation of
hyperfine induced E1 transition probability for $2s2p~^3P_0, ~^3P_2
\rightarrow 2s^2~^1S_0$ Eq. (11) is can be simplified to
\begin{eqnarray}
A = \frac{4 {\omega}^3}{9 c^3} \left| \sum_{S=0,1} h_{S} \langle
2s^2~^1S_0  \|  Q^{(1)}  \| 2s2p~^{(2S+1)}P_1 \rangle
\right|^2,
\end{eqnarray}
The differences for these two transitions concentrate on different
hyperfine mixing coefficient $h_S$ and transition energy $\omega$.
As can be seen from Eq. (16), there exist interference effects caused
by the $^3P_1$ and $^1P_1$ two transition amplitudes.

The accuracy of the electronic matrix elements involved in above
formula was determined by wavefunction of electronic part. The main
uncertainty in calculation of the wavefunctions comes from
electron correlation effects, especially for low charged ions. For
instance C$^{2+}$ ions, those matrix elements are fair sensitive to
electronic correlation effects \cite{Ynnerman, Joensson2, Chen}.
Therefore, it must be considered very carfully. As a starting point
SCF calculations were done for the configurations in the even and
odd state complex. In the calculations the wavefunctions of
$1s^22s^2~^1S_0$, $1s^22s2p ~^3P_{0,1,2}$ and $1s^22s2p~^1P_1$ were
determined in extended optimal level (EOL) calculations
\cite{Dyall}, respectively. These calculation were followed by
calculations with expansions including configuration state functions
obtained by single(S) and double(D)-excitations from, respectively,
the studied even and odd state reference configurations to active
sets of orbitals $n \le 5$. While for C$^{2+}$ the active set was
expanded to $n \le 7$, and for N$^{3+}$ and F$^{5+}$ to $n \le 6$ in
order to obtain satisfied results. The active sets were
systematically increased allowing computed properties to be
monitored. Due to stability problems in the relativistic SCF
procedure only the outermost layers of orbitals could be optimized
each time. The SCF calculations were followed by CI calculations in
which part of core-valence and core-core correlations, the frequency
independent Breit interaction and QED effects was included. Finally,
the hyperfine induced transition probability can be obtained by
above mentioned theoretical method using HFST package \cite{HFST}.

\subsection{Hyperfine quenching rate}
In Table 1 we listed our calculated results for hyperfine induced
transition probability as well as other theoretical and experimental
value\cite{Brage, Cheng, Marques, Schippers}. Nuclear parameters of
concerned isotopes for Be-like ions were taken from \cite{Stone}. As
can be seen from Table 1, results of Marques {\it et al.} obviously
deviate from others because two problems. One is that they neglected
the contribution of $^1P_1$ to hyperfine quenching rate of $^3P_0$,
and the other is that a ratio factor in transition rate was omitted
in their method \cite{Cheng}. The present calculational results
displayed in the forth column in Table 1 are in agreement with
others for $Z \le 30$, but not for ions with $Z>30$. The reason for
this is we neglected high order Breit interactions and QED effects
in electronic wavefunction calculations, which lead to quite large
errors in transition energy for high-Z ions. Due to limits of present
method, we have not intrinsically solved this problem. Hence
experimental \cite{NIST} and some of other accurate theoretical
transition energies \cite{Cheng} were used to correct the hyperfine
induced transition probability of $^3P_0$ levle. Those corrected ones were
presented in the fifth column of Table 1. It can be found that the
consistence becomes obviously better with Brage {\it et al.} and
Cheng {\it et al.} calculational value. While for Be-like $^{103}$Rh
ion the two order of magnitude difference is because different
magnetic dipole moment was used. Based on this, we used the same
method to correct other transition rates in next computations as
well.

In Table 2 we presented hyperfine quenching rate of $^3P_2$ in
connection with corresponding transition energies used to correct
those rates. Since angular momentum $J$ of $^3P_2$ state unequal
zero, it splits into several hyperfine levels, labeled by total
angular momentum $F$. For individual hyperfine level satisfying
select rule of electric dipole can occur hyperfine induced E1
transition, and this transition rate is dependent on angular number
$F$ and nuclear parameters.

Due to lack of other results about hyperfine induced $2s2p~^3P_2 \rightarrow 2s^2~^1S_0$ transition probability we can not make comparison.
In order to confirm the rates is
reliable, therefore, we further evaluated $2s2p~^3P_2 \rightarrow 2s^2~^1S_0$
M2 and $2s2p~^3P_2 \rightarrow 2s2p~^3P_1$ M1 transition
probabilities using our calculated line strength and experimental transition energies.
These results as well as other theoretical values were displayed in Table 3.
As can be seen from this Table, the consistence among these results is quite
good. It is indicated that our calculational hyperfine quenching rates of $^3P_2$ level are credible.

\subsection{Interference effects in hyperfine quenching}
Brage {\it et al.} and Cheng {\it et al.} have dictated that
interference effects strongly affect hyperfine quenching rate of
$^3P_0$ \cite{Brage, Cheng}. From Eq.(16), we know that this effect
occur in hyperfine quenching of $^3P_2$ as well. To show clearly
this phenomena, independent nuclear parameter of the transition
amplitudes contributed from $^3P_1$ and $^1P_1$ were plotted in Fig.
1. As can be seen from this picture, the interference effects work
within a wide range of atomic number due to non-monotone change of
transition amplitude contributed by $^1P_1$. It is interesting that
the trend of transition amplitude with Z is similar between those
two hyperfine quenching. An obvious difference is that the
transition amplitude of $^1P_1$ is dominant in $2s2p~^3P_2
\rightarrow 2s^2~^1S_0$ E1 transition while $^3P_1$ for $2s2p~^3P_0
\rightarrow 2s^2~^1S_0$ E1 transition.

In order to reveal characteristic of the interference effects in
such hyperfine induced $2s2p~^3P_0, ~^3P_2 \rightarrow 2s^2~^1S_0$
transition of Be-like ions, it is convenience to define a function,
$R^{el}$, that proportion to the ratio between the two transition
amplitudes. For example, for $2s2p~^3P_0 \rightarrow 2s^2~^1S_0$
transition,
\begin{eqnarray}
R^{el}(^3P_0) &=& |\frac{h_1\langle 2s^2\ ^1S_0 || Q^{(1)}||
2s2p\ ^3P_1 \rangle}{h_0\langle 2s^2\ ^1S_0 || Q^{(1)}|| 2s2p\
^1P_1 \rangle}|-1.
\end{eqnarray}
According to the formula, the closer $R^{el}$ is to 0, the stronger
interference effect is. Therefore, it clearly show the extent of
interference effects. The trend of $R^{el}$ with Z for $^3P_0$ and
$^3P_2$ is plotted in Fig. 2, respectively. It worth noting from
this picture that the interference effects in these two hyperfine
quenching do not change monotonically and there exist minimum value
for $R^{el}(^3P_0)$ near $Z=7$ and near $Z=9$ for $R^{el}(^3P_2)$.
Hence, the strongest interference effect occurs near $Z=7$ for
$2s2p~^3P_0 \rightarrow 2s^2~^1S_0$ E1 transition, and near $Z=9$
for $2s2p~^3P_2 \rightarrow 2s^2~^1S_0$ E1 transition.

\section{Conclusion}
In conclusion, we have calculated the hyperfine induced $2s2p~^3P_0,
~^3P_2 \rightarrow 2s^2~^1S_0$ E1 transition probability of Be-like
ions using grasp2K based on multiconfiguration Dirac-Fock method and
HFST packages. The interference effects resulted from $^3P_1$ and
$^1P_1$ perturbative states in those two hyperfine quenching of
$2s2p~^3P_0, ~^3P_2 \rightarrow 2s^2~^1S_0$ were studied in detail.
It worth noting that the
trends of interference effects with atomic number $Z$ in such two
transitions are not monotone. The strongest interference effect
occurs near $Z=7$ for $2s2p~^3P_0 \rightarrow 2s^2~^1S_0$ E1
transition, and near $Z=9$ for $2s2p~^3P_2 \rightarrow 2s^2~^1S_0$
E1 transition.

\section{Acknowledgments}
We would like to thank Prof. Per J\"onsson
and Prof. Gediminas Gaigalas for their helpful discussions.
This work has been supported by the National Nature Science
Foundation of China (Grant No. 10774122, 10876028), the specialized
Research Fund for the Doctoral Program of Higher Education of China
(Grant No. 20070736001) and the Foundation of Northwest Normal
University (NWNU-KJCXGC-03-21).

\section{Figure caption}
Fig. 1. Transition amplitude of reduced hyperfine induced transition in
a.u.. Left: $2s2p~^3P_0 \rightarrow 2s^2~^1S_0$ transition; Right:
$2s2p~^3P_2 \rightarrow 2s^2~^1S_0$ transition.

Fig. 2. The trend of $R^{el}$ for $^3P_0$ and $^3P_2$ level with
atomic number Z. Two blue arrows label the positions where inference
effects are the strongest for the hyperfine induced $2s2p~^3P_0 \rightarrow 2s^2~^1S_0$
and $2s2p~^3P_2 \rightarrow 2s^2~^1S_0$ transition, respectively.

\clearpage

\begin{table}[h]
{\footnotesize \caption{\footnotesize Hyperfine induced $2s2p~^3P_0
\rightarrow 2s^2~^1S_0$ transition probability in s$^{-1}$ and
corresponding transition energy in cm$^{-1}$. The calculational
results were compared with other theoretical and experimental value.
uncorr. means the hyperfine induced transition probabilities was
computed using present calculational transition energy, while the
corr. used the ones from NIST database \cite{NIST}.}
\begin{center}
\begin{tabular}{lccclcccccccccccccc}
\hline \hline
&\multicolumn{2}{c}{Transition energy}&&\multicolumn{2}{c}{This work}&&&&&\\
\cline{2-3}\cline{5-6}ions & This work & NIST \cite{NIST} && uncorr. & corr. &Ref.\cite{Brage} & Ref.\cite{Cheng} & Ref.\cite{Marques} & Expt.    \\
\hline
$^{13}$C   &52248  & 52367              &&8.28[-4]  & 8.33[-4]& 9.04[-4]&  8.223[-4] &2.00[-4] &         \\
$^{14}$N   &67251  & 67209              &&4.40[-4]  & 4.39[-4]& 4.92[-4]& 4.40[-4]   &1.28[-4] & 4[-4]$\pm$1.32$^a$        \\
$^{19}$F   &96666  & 96590              &&1.17[-1]  & 1.17[-1]&         &  1.208[-1] &3.60[-2] &         \\
$^{28}$Si  &169054 & 169802             &&5.89[-2]  & 5.97[-2]&6.08[-2] &  6.011[-2] &2.16[-2] &         \\
$^{39}$Ar  &228716 & 228674             &&8.28[-1]  & 8.27[-1]&         &            &         &         \\
$^{47}$Ti  &289562 & 288190             &&6.80[-1]  & 6.71[-1]&         &  6.727[-1] &3.56[-1] &5.6[-1]$^b$ \\
$^{57}$Fe  &352029 & 348180             &&4.98[-2]  & 4.82[-2]&5.45[-2] &  4.783[-2] &3.27[-2] &         \\
$^{67}$Zn  &416600 & 409827$^{\dagger}$ &&5.00      & 4.76    &         &  4.732     &4.13     &         \\
$^{85}$Rb  &537174 & 523000             &&43.3      & 39.94   &         &  39.35     &48.17    &         \\
$^{103}$Rh &693209 & 661772$^{\dagger}$ &&147.1     & 128.0   &         &  1.262     &1.91     &         \\
$^{131}$Xe &903919 & 843105$^{\dagger}$ &&199.0     & 161.5   &         &  158.1     &262.67   &         \\
\hline
\end{tabular}
\end{center}
$^{\dagger}$ Cheng {\it et al.} \cite{Cheng} \\
$^a$ Brage {\it et al.} \cite{Brage2}\\
$^b$ Schippers {\it et al.} \cite{Schippers}}
\end{table}

\clearpage

\begin{table}[h]
{\footnotesize  \caption{\footnotesize Hyperfine induced $2s2p~^3P_2
\rightarrow 2s^2~^1S_0$ E1 transition probability A in s$^{-1}$
associated with corresponding reduced transition rate A$^{el}$ in
s$^{-1}$ and transition energy $\Delta E$ from NIST database \cite{NIST} in
cm$^{-1}$.}
\begin{center}
\begin{tabular}{lccccccccccccccccccccccc}
\hline \hline
ions  & $\Delta E$ &  A$^{el}$ & F & A &  &  ions &  $\Delta E$ &  A$^{el}$    & F &  A \\
\hline
$^{13}$C  & 52447 & 9.87[-4]  & 3/2&  7.30[-4]  &   & $^{57}$Fe   & 471780     & 1.24      & 3/2&  1.52[-1]                 \\
          &       &           & 5/2&     0      &   &             &            &           & 5/2&    0                  \\
$^{14}$N  & 67412 & 2.52[-4]  & 1  &  2.08[-4]  &   & $^{67}$Zn   & 640470     & 4.40      & 1/2  &        0                  \\
          &       &           & 2  &  3.69[-4]  &   &             &            &           & 3/2  &      8.81                 \\
          &       &           & 3  &     0      &   &             &            &           & 5/2  &      1.75[1]              \\
$^{19}$F  & 97437 & 1.21[-3]  & 3/2&  1.26[-1]  &   &             &            &           & 7/2  &      1.79[1]              \\
          &       &           & 5/2&     0      &   &             &            &           & 9/2  &         0                  \\
$^{28}$Si & 177318& 1.70[-2]  & 3/2&  7.68[-2]  &   &  $^{85}$Rb  & 1094800    & 3.83[1]   & 1/2  &        0          \\
          &       &           & 5/2&    0       &   &             &            &           & 3/2  &      1.85[2]                   \\
$^{39}$Ar & 252683& 8.28[-2]  & 3/2&    0       &   &             &            &           & 5/2  &      3.65[2]             \\
          &       &           & 5/2&  5.83[-1]  &   &             &            &           &7/2  &      3.71[2]               \\
          &       &           & 7/2&  1.03      &   &             &            &           &9/2  &         0                      \\
          &       &           & 9/2&  9.76[-1]  &   &  $^{103}$Rh & 2310547$^{\dagger}$& 4.08[2]   &3/2   &      4.79[2]        \\
          &       &           & 11/&    0       &   &             &            &           &5/2  &         0                       \\
$^{47}$Ti & 347420& 3.34[-1]  & 1/2&    0       &   &  $^{131}$Xe & 3785850    & 4.96[3]   &1/2   &      3.95[3]     \\
          &       &           & 3/2&  4.82[-1]  &   &             &            &           &3/2  &      1.30[4]                      \\
          &       &           & 5/2&  1.03      &   &             &            &           &3/2  &      1.30[4]  \\
          &       &           & 7/2&  1.17      &   &             &            &           &     &                \\
          &       &           & 9/2&    0       &   &             &            &           &     &                    \\

\hline
\end{tabular}
\end{center}
$^{\dagger}$ Cheng {\it et al.} \cite{Cheng}}
\end{table}

\clearpage

\begin{sidewaystable}[t]
{\scriptsize \caption{\scriptsize $2s2p~^3P_2 \rightarrow
2s^2~^1S_0$ M2 and $2s2p~^3P_2 \rightarrow 2s2p~^3P_1$ M1 transition
probabilities of Be-like ions in s$^{-1}$ connecting with corresponding transition
energy from NIST database \cite{NIST} in cm$^{-1}$. The calculational results
were compared with other theoretical and experimental value}
\begin{center}
\begin{tabular}{lcccccccccccccccccccc}
\hline \hline
&  \multicolumn{4}{c}{M2 ($~^3P_2 - ~^1S_0$) } && \multicolumn{5}{c}{M1($~^3P_2 - ~^3P_1$)} \\
\cline{2-5}\cline{7-11} ions & Transition energy &  This work &   Ref. \cite{Glass}  & Ref. \cite{Majumder} && Transition energy &This work & Ref. \cite{Kingston} &Ref. \cite{Tupitsyn} & Ref. \cite{Fischer2} \\
\hline
$^{13}$C   & 52447               &5.13[-3]   & 5.190[-3] &  5.176[-3]  && 56                  & 2.37[-6] & 2.34[-6] &            &  2.446[-6]             \\
$^{14}$N   & 67416               &1.14[-2]   & 1.154[-2] &  1.147[-2]  && 144                 & 4.03[-5] & 3.93[-5] &            &  4.070[-5]             \\
$^{19}$F   & 97437               &3.64[-2]   & 3.678[-2] &  3.633[-2]  && 587                 & 2.73[-3] &          &            &                          \\
$^{28}$Si  & 177318              &2.42[-1]   & 2.431[-1] &  2.410[-1]  && 5174                & 1.87     & 1.83     &            &                            \\
$^{39}$Ar  & 252683              &7.89[-1]   & 7.904[-1] &  7.858[-1]  && 16820               & 6.39[1]  & 6.41[1]  & 6.417[1]   &                             \\
$^{47}$Ti  & 347240              &2.44       &           &  2.4234     && 42620               & 1.03[3]  &          & 1.0369[3]  &                              \\
$^{57}$Fe  & 471780              &7.68       & 7.652     &  7.6459     && 92655               & 1.04[4]  & 1.11[4]  &            &                             \\
$^{67}$Zn  & 640470              &2.54[1]    &           &  2.5289[1]  && 180855              & 7.52[4]  &          &            &                             \\
$^{85}$Rb  & 1094800             &2.28[2]    &           &             && 480900              & 1.31[6]  &          &            &                              \\
$^{103}$Rh & 1996313$^{\dagger}$ &2.90[3]    &           &  2.8789[3]  && 1198404$^{\dagger}$ & 1.86[7]  &          &            &                             \\
$^{131}$Xe & 3785850             &4.62[4]    &           &             && 2758850             & 2.10[8]  &          &            &                             \\
\hline
\end{tabular}
\end{center}
$^{\dagger}$ Cheng {\it et al.} \cite{Cheng}}
\end{sidewaystable}

\clearpage

\begin{figure}[h]
\centering
\includegraphics[scale=0.7]{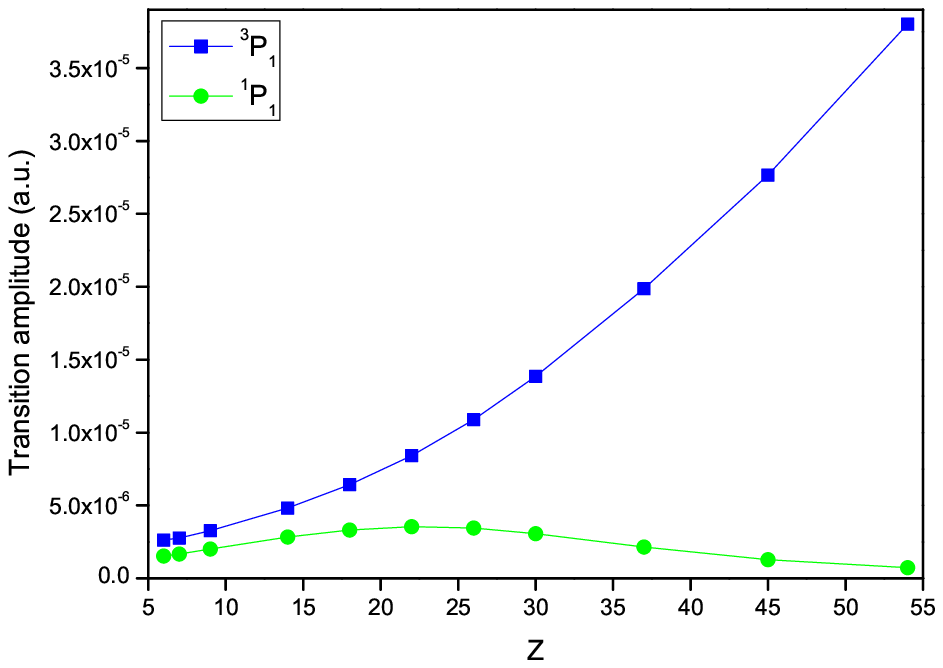}
\includegraphics[scale=0.7]{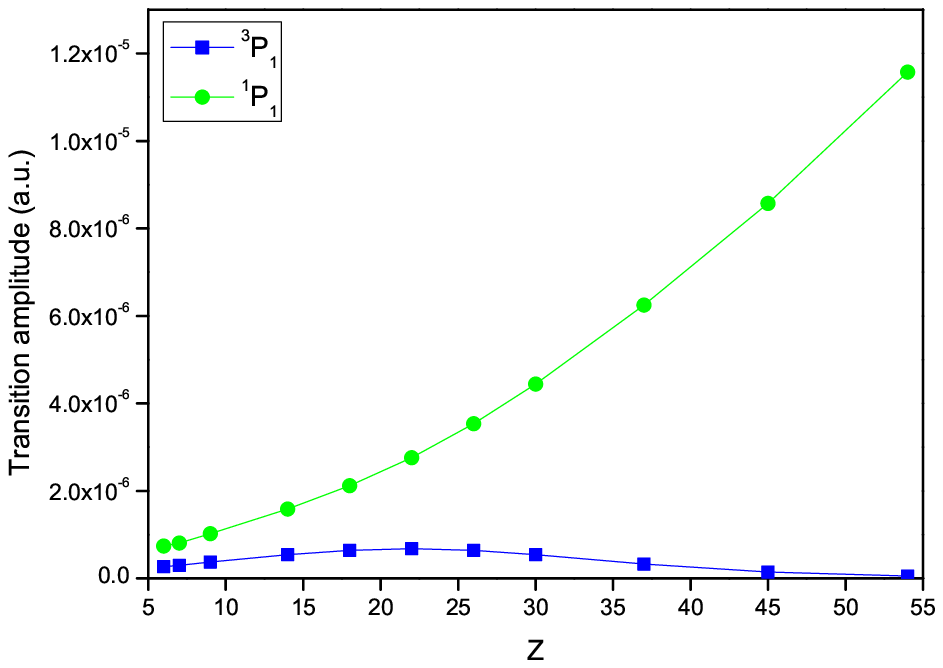}
\end{figure}

\clearpage

\begin{figure}[t]
\centering
\includegraphics[scale=0.8]{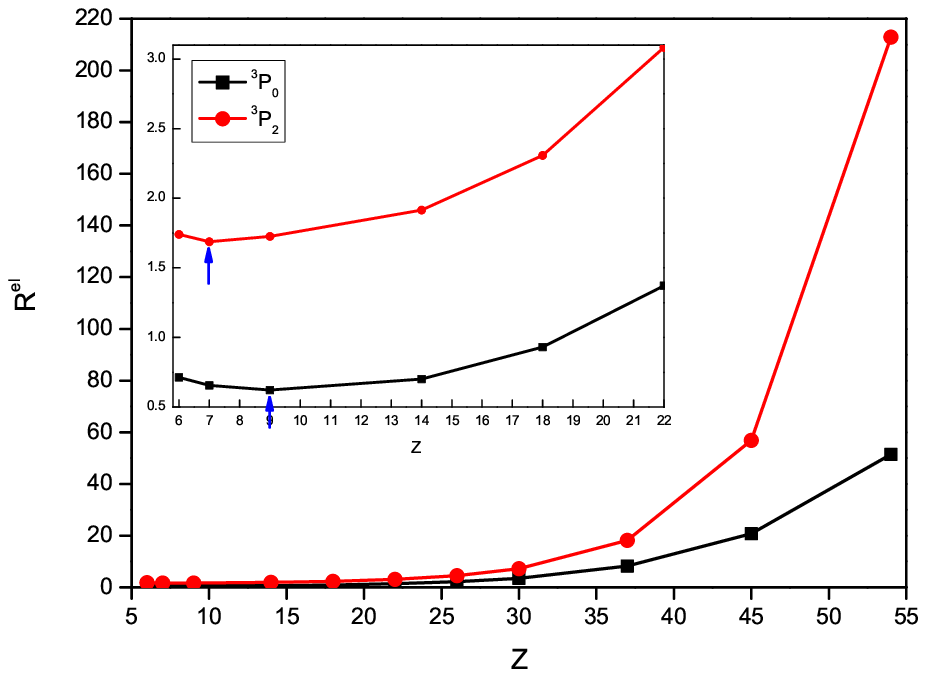}
\end{figure}

\end{document}